\documentclass[epj,final]{svjour}
\usepackage{graphics,epsfig}

\newcommand{\kin}{k^{\rm in}}
\newcommand{\fluck}{\left< k^2 \right>}
\begin{document}
\title{Disease Spreading in Structured Scale-Free Networks}
\author{Yamir Moreno\inst{1}\thanks{e-mail: yamir@ictp.trieste.it} \and Alexei V\'azquez\inst{2}
} 
%
%
\institute{The Abdus Salam International Centre for Theoretical
Physics, P.O. Box 586, 34100 Trieste, Italy \and International School
for Advanced Studies SISSA/ISAS, via Beirut 4, 34014 Trieste, Italy}
\date{Received: \today / Revised version: }
%
\abstract{We study the spreading of a disease on top of structured scale-free
networks recently introduced. By means of numerical simulations we
analyze the SIS and the SIR models. Our results show that when the
connectivity fluctuations of the network are unbounded whether the
epidemic threshold exists strongly depends on the initial density of
infected individuals and the type of epidemiological model
considered. Analytical arguments are provided in order to account for
the observed behavior. We conclude that the peculiar topological
features of this network and the absence of small-world properties
determine the dynamics of epidemic spreading.}

\PACS{{89.75.-k}{Complex Systems} \and {89.75.Fb}{Structures and
      organization in complex systems} \and {05.70.Jk}{Critical point
      phenomena}} 
%
\maketitle


\section{Introduction}

During the last years, there has been a burst of activity in the study
of complex networks \cite{barabasi02,dorogorev}. It has been shown
that many social and natural systems
\cite{falou99,alexei,www99,strog01,montoya02,wagner01,jeong01,spsk,vazquez}
can be represented as a graph where nodes represent individuals or
agents and links stand for the physical interactions among
them. Surprisingly, many of these networks share some important
topological features such as small-world (SW) properties
\cite{watts98} and scale-free (SF) degree distributions
\cite{barab99}, where the degree or connectivity $k$ of a node is the number of
neighbors to which it is linked. Networks displaying scale-free degree
distributions are very interesting not only for their relative
abundance in Nature but also because of their peculiar statistical
properties. In particular, the unbounded fluctuations of the
connectivity distribution $P(k)\sim k^{-\gamma}$ ($\gamma \le 3$) seem
to be a blueprint of all real-world networks studied so far
\cite{barabasi02,dorogorev}.

The SW and SF properties have considerable impact on the
processes running on top of complex networks. The effects of the
complex topological features of SF networks on the dynamics of
epidemic disease spreading is perhaps one of the most interesting
outcomes in the study of complex networks. Specifically, the
understanding of spreading phenomena in these networks can shed light
on a large number of practical problems, ranging from computer virus
infections \cite{newvir,white} to epidemiology
\cite{murray,anders}. For instance, it has been recognized only
recently \cite{pv01a,pv01b} that random SF networks are completely
prone to epidemic spreading allowing the onset of large epidemics
whatever the spreading rate of the infection. This radical change with
respect to the way a disease is spread in a regular structure is
rooted in the diverging connectivity fluctuations of SF networks with
$\gamma \le 3$. Similarly, random SF networks have been shown to
exhibit extremely robustness to random damages
\cite{barabasi00,newman00,havlin01}.

On the other hand, real networks are also characterized by degree
correlations that might play a fundamental role in the functional
properties of networked systems and on processes running on correlated
networks. The study of networks with degree correlations and the
extension of previous results obtained for random networks to
correlated networks is quite recent
\cite{berg02,marian1,marian3,n02a,vw02,cond182}. One of the models
aimed to account for degree correlations was proposed recently by
Klemm and Egu\'{\i}luz \cite{klemm02} (henceforth referred to as
deactivation model) where nodes can be deactivated with probability
inversely proportional to their connectivity. The model turns out to
be more rich than initially thought with a power law connectivity
distribution $P(k)\sim k^{-\gamma}$ but with $\gamma$ laying in a
quite wide interval $2 \le \gamma \le 4$ \cite{cond183}. Additionally,
it has been shown that the topology of the network is essentially
linear being the graph a collection of stars of diverse degree
connected as a chain by a number of {\em local} links, i.e., the graph
lacks SW properties. While a threshold may exist \cite{structured} due
to the peculiar topological properties of the network that makes the
epidemic spreading to be dominated by the diffusion of the disease on
a linear chain \cite{cond183}, the star-like graphs connected as a
chain might also lead to new effects in the dynamics of epidemic
spreading.

In this paper, we study in detail by large scale numerical simulations
two paradigmatic epidemiological models, namely, the
Susceptible-Infected-Susceptible (SIS) and the
Susceptible-Infected-Removed (SIR) models on top of networks generated
using the deactivation
model. We found that the existence
or not of an epidemic threshold for $\gamma \le 3$ depends on the
initial density of infected individuals while for values of $\gamma >
3$ the epidemic threshold is recovered and it is universal, i.e., it
does not depend on the initial density of infected individuals. We
also provide analytical arguments in order to explain our numerical
findings. Moreover, we show that the linear character of the graph
determines the existence of a trivial threshold for the SIR model that
can be mapped to a bond percolation problem in one dimension.

\section{Deactivation model}

Let us first summarize the main properties of the deactivation model
introduced by Klemm and Egu{\'\i}luz \cite{klemm02}. The growing
dynamics of the network is defined in the following way: One starts
from a fully connected graph of $m$ nodes that are set active. At
successive time steps nodes are added one by one following the recipe:
$(i)$ A new node is 
connected to all active nodes in the network; $(ii)$ one of the active
nodes is selected for deactivation with probability
\begin{equation}
  p_d(\kin_i)=\frac{\left[\sum_{j\in{\cal A}}(a+\kin_j)^{-1}\right]^{-1}}{a+\kin_i};
\label{eq:1}
\end{equation}
where the sum in Eq.~(\ref{eq:1}) runs over the set of active nodes
${\cal A}$, $a$ is a model parameter, and $\kin_i$ denotes the
in-degree of the $i$-th node. Finally, $(iii)$ the new node just added
is set active.

Recently, it has been shown \cite{cond183} that most of the topological
properties (in particular, the degree distribution) of the network
generated using the above rules are very sensitive to the order in
which steps $(ii)$ and $(iii)$ are performed. Thus we shall
discriminate in the following two cases \cite{cond183}: 
\begin{itemize}
\item {\em Model A}: $(ii)$ is performed \textit{before} $(iii)$ .
\item {\em Model B}: $(iii)$ is performed \textit{after} $(ii)$.
\end{itemize}

The deactivation model is usually run taking $a=m$. This makes the
deactivation probability inversely proportional to the total
connectivity of the nodes $k=m+\kin$. The connectivity distribution
can be analytically obtained for the limiting cases of lowest
\cite{cond183} and largest $m$ \cite{structured} for each model
resulting in a power law that reads as,
\begin{equation}
P(k)\sim k^{-\gamma},
\label{eq:3}
\end{equation}
where the exact value of $\gamma$ depends on the model considered and
the value of $m$ such that,
\begin{itemize}
\item {\em model A} with $a=m$ $\quad \Longrightarrow \quad  3<\gamma\leq4$
\item {\em model B} with $a=m$ $\quad \Longrightarrow \quad  2\leq\gamma<3$.
\end{itemize}

Noticeable and relevant for the physical processes run on top of this
model, the dynamics of the deactivation model allows exponents
$\gamma$ of the connectivity distribution that can lead to an
unbounded second moment ($\fluck \rightarrow \infty$ for model B) or
to finite connectivity fluctuations ($\fluck$ is finite for model
A). More important, however, is the fact that the generated networks
lack SW properties. Small-world properties refer to the fact that in
many complex networks one can go from one node to any other node of
the system visiting a very small number of intermediate nodes. The
minimum number of such intermediate nodes for each pair $(i,j)$ of
nodes in the network is called the minimum path length between $i$ and
$j$, and the diameter of the network is defined as the largest among
the shortest paths between any two nodes in the network. For networks
displaying SW properties, the graph diameter grows, at most,
logarithmically with the system size $N$.

The networks generated with the deactivation model consist of a
collection of star-like graphs connected as a chain, resembling the
topology of a one dimensional lattice. In fact, irrespective of the
version considered, both the diameter and the average minimum path
distance scale linearly with the network size $N$ \cite{cond183}. In
particular, the mean square displacement of a random walker on the
deactivation model scales with time almost as for a one dimensional
lattice qualifying for a slightly subdiffusive behavior
\cite{cond183}. Hence, the existence of an epidemic threshold in the
deactivation model is not surprising. However, at the same time, the
peculiar characteristics of the graph and, in particular, the presence
of stars with very large connectivities introduces new and interesting
effects in the dynamics of disease spreading. As we shall discuss in
the following section, whether or not a finite prevalence sets in for
the SIS model would depend on the initial density of infected
individuals.


\section{The SIS Model}

In the standard SIS epidemiological model \ \cite{murray}, each node
of the network represents an individual and each link is the
connection along which the individuals interact and the epidemic can
be transmitted. In this model, the individuals can exist in two
possible states, namely, susceptible or healthy and
infected. Individuals removal due to death or acquired immunization is
not allowed and thus they stochastically move through the cycle
susceptible $\rightarrow$ infected $\rightarrow$ susceptible. The
disease transmission is described in an effective way. At each time
step, each susceptible node connected to one or more infected nodes
gets the infection with probability $\lambda$, while the infected
nodes are cured and become again susceptible at a rate $\delta$
(henceforth set to 1 without lack of generality). In the SIS model
whenever the epidemic pervades the system, it gets into an endemic
state with a stationary value for the density of infected nodes that
acts as the order parameter of the model.

Results on regular structures point out that there is a threshold
below which the system does not reach such a stationary state and the
infection dies out \cite{murray,pv01b}. Moreover, the value of the
prevalence at the stationary state and its own existence does not depend
on the initial density of infected individuals. One can start with an
initial density $\rho_0=1/N$ of infected individuals or from, let's
say, half of the lattice infected (in general, $\rho_0=const$): The
prevalence self-organizes into a stationary state in both cases. It is
worth noting that both situations are feasible in practice and thus
physically relevant. 

On the other hand, the behavior of the SIS model on random SF networks
is radically different \cite{pv01a,pv01b} when $\gamma \leq 3$. In
this case, there is no epidemic threshold in the thermodynamic limit
and the networks are completely prone to the spreading of the
disease. This drastic change of behavior is due to the diverging
connectivity fluctuations of SF networks. Moreover, it has also been
shown that for finite-size system an effective threshold is recovered,
but its value is significantly overestimated \cite{fspv}.

In order to study the SIS model in deactivation model networks, we
first generate networks using the algorithm described in the preceding
section for both settings of the model A and B with $a=m=3$. Then we
let the system evolve according to the SIS dynamics. Large scale
numerical simulations were performed in networks of size up to
$N=10^6$, averaging over at least 100 different realizations on at
least 10 different realizations of the network. Initially, a fraction
of nodes $\rho_0$ was infected and we let the system relax into the
steady state where the prevalence $\rho$ attains its stationary
value. Two initial conditions for the density of infected individuals
were considered. In the first case, we start from a single infected
individual, $\rho_0=1/N$, and set the observation time $t_m$ to be
$10^6$ time steps, that is, the state of each node is updated up to
$t_m$ times if there is at least one infected individual. Then, we
repeat the process but starting with homogeneous initial
conditions. This is achieved by infecting a finite fraction of the
network $\rho_0=const.$ and monitoring as before whether a stationary
state sets in with its corresponding prevalence for the same
observation time $t_m$. It is worth recalling that the prevalence in
the stationary state is computed as the average over all surviving
trials in both cases.

\begin{figure}
\centerline{\psfig{file=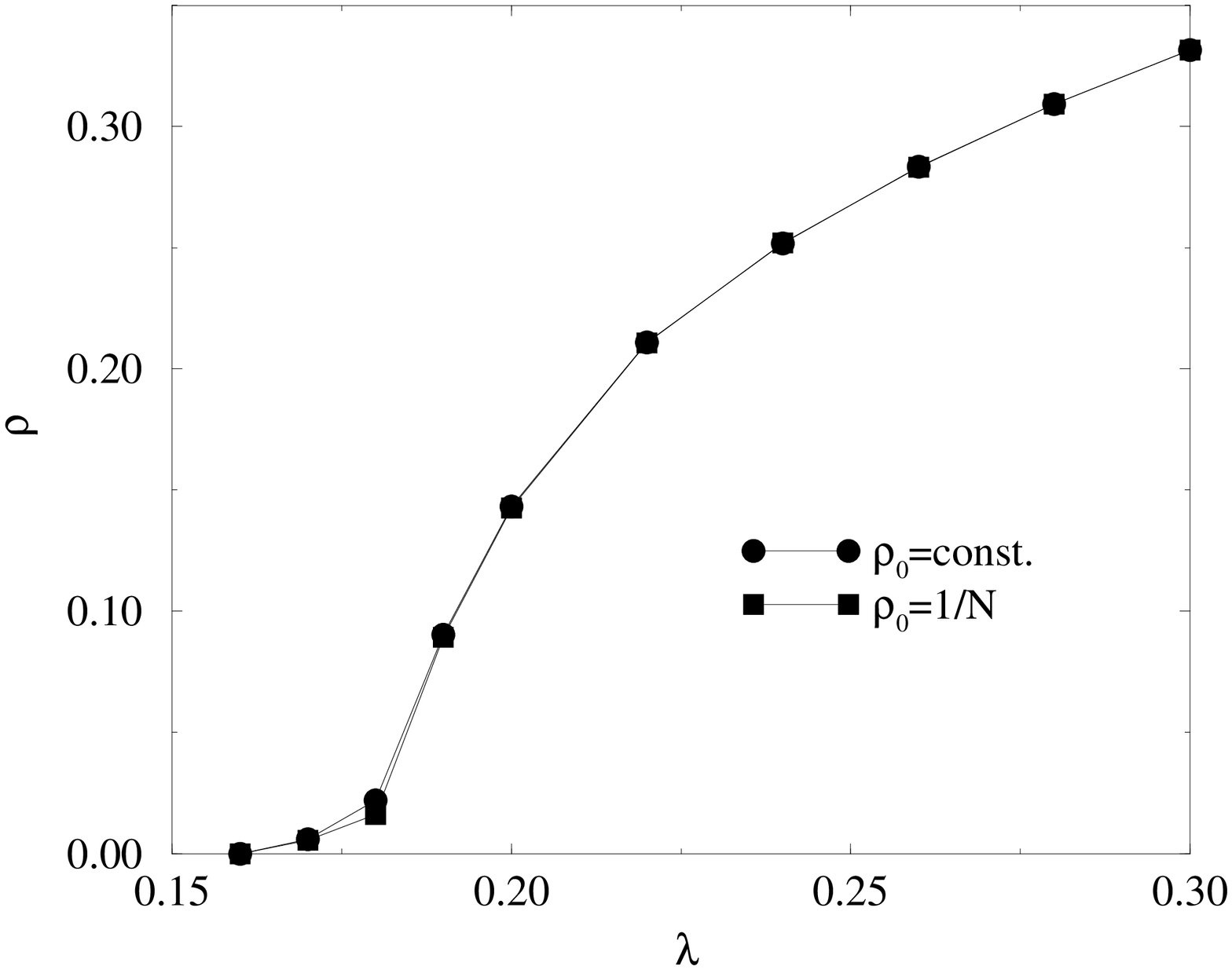,width=2.5in,angle=-90}}
\centerline{\psfig{file=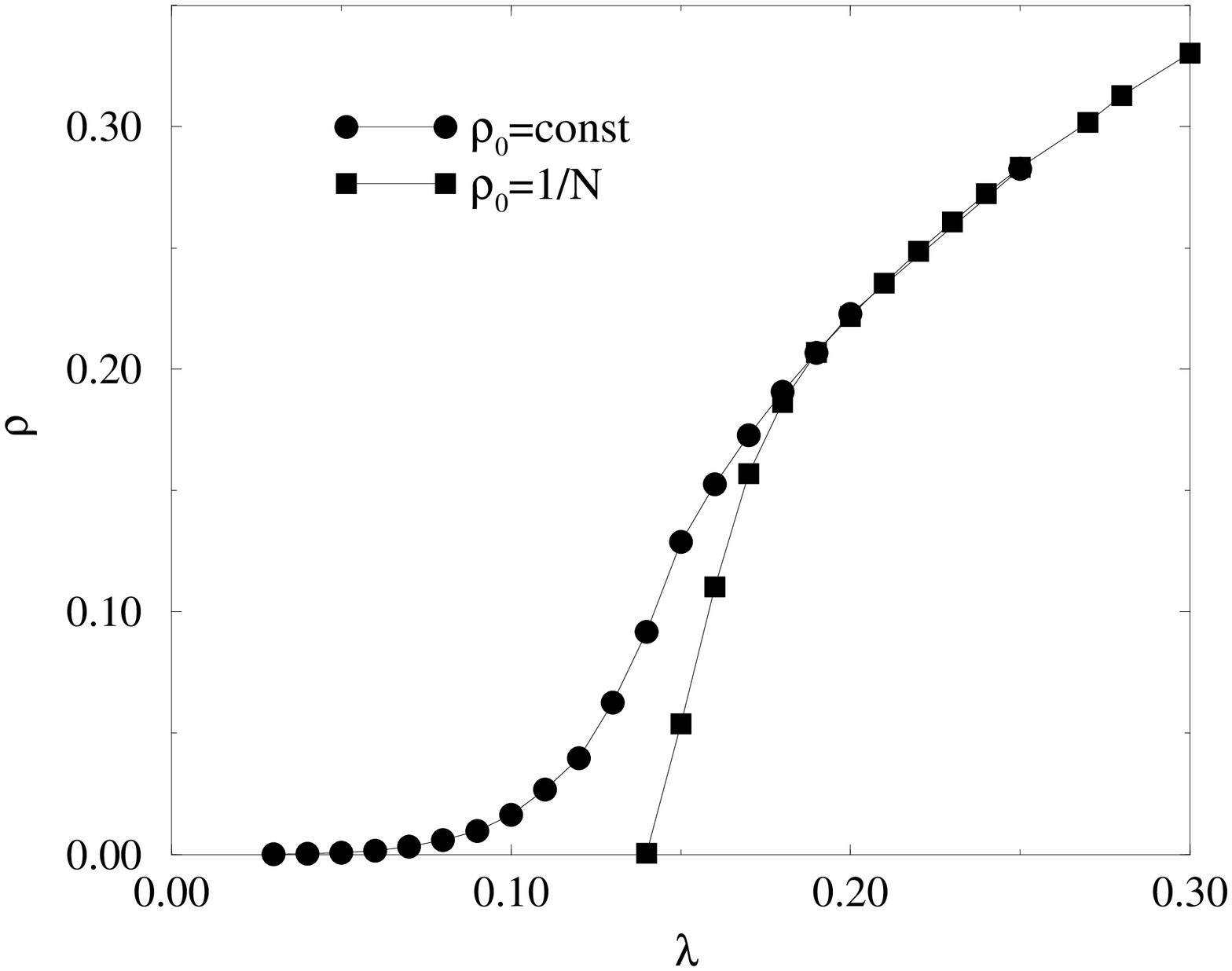,width=2.5in,angle=-90}}
\caption{ Phase diagram for the deactivation model A (top)  and B (bottom). The density of
infected nodes $\rho$ for two initial conditions is plotted as a
function of the spreading rate $\lambda$. The size of the system is
$N=10^6$ and the average connectivity was set to $\left< k\right>=6$
($m=3$). For model A the epidemic threshold is independent of $\rho_0$ and is
equal to $\lambda_c=0.16(1)$. For model B the epidemic
threshold depends on $\rho_0$. For an initial state with a single
infected individual $\lambda_c=0.14(1)$, whereas for homogeneous
initial conditions $\lambda_c \to 0$ in the thermodynamic limit.}
\label{fig2}
\end{figure}

The results obtained are depicted in Fig. \ref{fig2} where the steady
state prevalence has been drawn as a function of the spreading rate
$\lambda$. As expected for model A the epidemic threshold would appear
again for both initial conditions since in this case together with the
lack of SW properties and the linearity of the network, the second
moment $\fluck$ is finite with $3 < \gamma \le 4$. In this case, the
phase diagram is the same regardless of the initial density of
infected individuals $\rho_0$. It is worth noting that whenever an
epidemic threshold exists, its value seems to be the same (within
statistical errors) for both versions of the deactivation model,
depending only on the average connectivity $\langle k \rangle$. This
implies in its turn that the dynamics of the infection spreading is
mainly determined by the linear star-like structure of the graphs.

Two radically different behaviors can be noted for the deactivation
model B depending on the initial condition. In this case the exponent
of the power law decay of the degree distribution is in the interval
$2 \le \gamma < 3$. In particular a best fit analysis for networks of
size $N=10^7$ yields an exponent $\gamma=2.65 \pm 0.05$ for
$m=3$. When the system evolves from a single infected individual, an
epidemic threshold $\lambda_c$ appears, below which the epidemic
cannot pervade the network. In this case the fact that in the
thermodynamic limit the connectivity fluctuations are diverging is
canceled by the effects introduced by the linear topological nature of
the network and the lack of SW properties. As advanced in
Ref. \cite{cond183}, the SIS dynamics can be reduced in a coarse
grained picture to the diffusion of the disease on a linear
chain. Thus, the classical picture for regular structures with an
epidemic threshold is recovered.

\begin{figure}
\centerline{\psfig{file=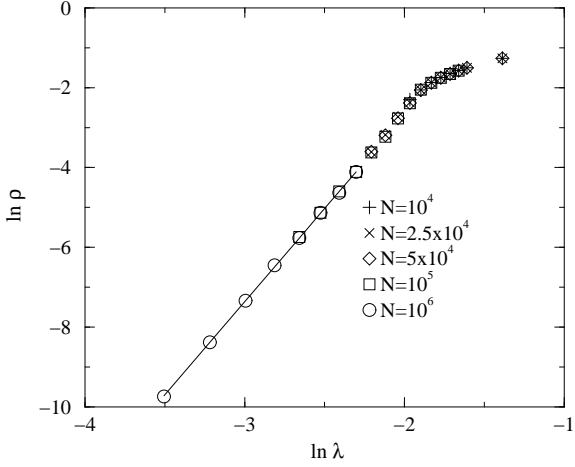,width=2.5in,angle=-90}}
\caption{ Prevalence $\rho$ for the SIS model in scale-free correlated
networks generated using deactivation model B. The average
connectivity of the network is in this case $\left< k\right>=6$
($m=3$). The straight line is a fit to the form $\rho(\lambda)\sim
\lambda^{\beta}$, with $\beta\approx 4.5$.}
\label{fig3}
\end{figure}

For homogeneous initial conditions, the phase diagram is completely
different. In this case, the stars with high connectivity can let the
epidemic survive and a final stationary state is reached by a long
power-law decay. In order to rule out the presence of finite size
effects hiding an abrupt transition we have checked the behavior of
the stationary prevalence for several system sizes. Note that larger
system sizes are needed in order to correctly depict the prevalence
behavior for $\lambda \ll 1$, since finite size networks induce finite
size corrections to the zero threshold \cite{pv01a,pv01b}. The results
obtained are plotted in Fig.\ \ref{fig3}. The straight line is a fit
to the form $\rho(\lambda)\sim \lambda^{\beta}$, with $\beta\approx
4.5$, showing that the infection prevalence is assuming a finite
stationary value for all values $\lambda > 0$. This numerical evidence
confirms the previous picture obtained for uncorrelated scale-free
networks, namely, the absence of an epidemic threshold in the
thermodynamic limit. It is interesting to recall that using a
dynamical mean field approach for uncorrelated networks with
scale-free connectivity distributions given by $P(k)\sim k^{-\gamma}$,
the same functional form $\rho(\lambda)\sim \lambda^{\beta}$ is
predicted when $2< \gamma < 3$, although in this case satisfying the
relation $\beta=1/(3-\gamma)$ \cite{pv01b}. Hence, it seems that
correlations, while preserving the general functional form, make the
prevalence to decay more faster than it does in the uncorrelated
case. This in turn may explain why in finite systems the epidemic
threshold is smaller than its counterpart in uncorrelated networks, a
fact also observed in random network models \cite{marian1}. Below, we
provide an analytical argument that help understand the origin of
these differences by analyzing the SIS model on an ensemble of
disconnected stars.

\subsection{SIS model on an ensemble of disconnected stars}

Let us study a graph made up of an ensemble of disconnected
stars. To be more precise, we consider a graph where all nodes
with degree $k>1$ are connected to nodes with degree $k=1$.
If $P(k)$ ($k\geq1$) is the degree distribution then
$P(1)=\sum_{k>1}kP(k)$. Since the stars are isolated the stationary
prevalence is given by
\begin{equation}
\rho = \sum_{k>1} P(k) n(k) v(k)\ ,
\label{rho_all}
\end{equation}
where $n(k)$ is the stationary number of infected nodes on a star of
degree $k$ and $v_k$ is the probability that at $t=0$ there was at
least one infected node in a star with $k$ leaves.

Now, to compute $n(k)$ we focus on the SIS dynamics on a single
star. The spreading takes place from the central node to the leaves,
then from the leaves to the central node and so on. For stars with a
large number of leaves ($k\gg1$) we can approximate the average number
of infected nodes by the number of infected leaves. Hence, the
number of infected nodes at step $t+1$ is
\begin{equation}
n(k,t+1) = \left[ 1- (1-\lambda)^{n(k,t)} \right] \lambda k\ .
\label{n_t}
\end{equation}
The first factor in the right hand side is the probability that the
central node gets infected, receiving the disease from at least one
leaf. The second factor is the number of infected leaves given that
the center was infected. This linear map has always the trivial
solution $n(k,t)=0$. To investigate its stability we assume
$n(k,t)\ll1$ resulting
\begin{equation}
n(k,t+1) = \lambda k\ln\frac{1}{1-\lambda} n(k,t)\ .
\label{n_t_1}
\end{equation}
Thus, there is a critical star size
\begin{equation}
k_c(\lambda) = \left( \lambda \ln\frac{1}{1-\lambda} \right)^{-1}\ ,
\label{d_c}
\end{equation}
such that for $k<k_c$ the prefactor in the right hand side of
Eq. (\ref{n_t_1}) is smaller than 1 and, therefore,
$n(k)=0$. On the contrary, for $k>k_c$ there is an exponential
growth of $n(k,t)$ indicating that the solution $n(k,t)=0$ is not
stable, {\em i.e.} $n(k)>0$. Moreover, the critical degree
depends on $\lambda$, with the limiting cases $k_c(0)=\infty$ and
$k_c(1)=0$.

Going back to Eq. (\ref{rho_all}) we obtain that
\begin{equation}
\rho = \sum_{k>k_c} P(k) n(k) v(k)\ .
\label{rho_all_final} 
\end{equation}
Let us now distinguish between two different initial conditions
considered before. In one case we start from only one infected
node. The probability that one node of a star of $k$ leaves is
infected is
\begin{equation}
v^{(1)}(k)= \frac{k+1}{N}\ .
\label{v_1}
\end{equation}
The substitution of this expression in Eq. (\ref{rho_all_final}) yields
\begin{equation}
\rho^{(1)}=\sum_{k>k_c} P(k) n(k)\frac{(k+1)}{N}\sim\frac{\fluck}{N}\ .
\label{rho_1}
\end{equation}
Now, for $2<\gamma<3$ the second moment grows at most as $\fluck={\cal
O}(N^{3-\gamma})$.  Hence, in the thermodynamic limit the prevalence
is asymptotically zero. However, a different result is obtained if we
infect a finite fraction $\rho_0$ of the nodes. In this case the
probability that at least one node of a star with $k$ leaves is
infected is given by
\begin{equation}
v^{(2)}(k) = 1-(1-\rho_0)^{k+1}\geq\rho_0\ .
\label{v_2}
\end{equation}
From this expression and Eq. (\ref{rho_all_final}) it follows that
\begin{equation}
\rho^{(2)} \geq \rho_0 \sum_{k>k_c} P(k) n(k)\ .
\label{rho_2}   
\end{equation}
Thus, there is a finite prevalence in the stationary state. 

These two opposite scenarios are the same observed in the numerical
simulations of the preceding section. When there is a finite fraction
of infected nodes at $t=0$, a finite prevalence is obtained. On the
contrary, when the initial infection is concentrated in only one
node, the stationary state reflects the topological nature of the
graph. For the ensemble of disconnected stars we obtain a stationary
state with no infected node, corresponding to the SIS model in a
zero-dimensional system. On the other hand, for the deactivation
model, we get a phase transition at a finite infection rate as
expected for a one-dimensional system \cite{marro99}.

\section{SIR model}

The SIR model assumes that individuals can exist in three classes:
susceptible, infected and removed. The main difference with the SIS
model is that once an individual gets infected it is removed and,
therefore, it can not catch the infection again. If we start from
a single infected node the size of the outbreak of the disease is
given by the number of nodes that can be reached assuming that each
link is occupied with a probability $\lambda$. Thus, the SIR model is
equivalent to a bond percolation problem with bond occupation
probability $\lambda$ \cite{newman02a}. Moreover, the size of the
outbreak is just the size of the giant component.

\begin{figure}[t]
\begin{center}
\includegraphics[width=3in]{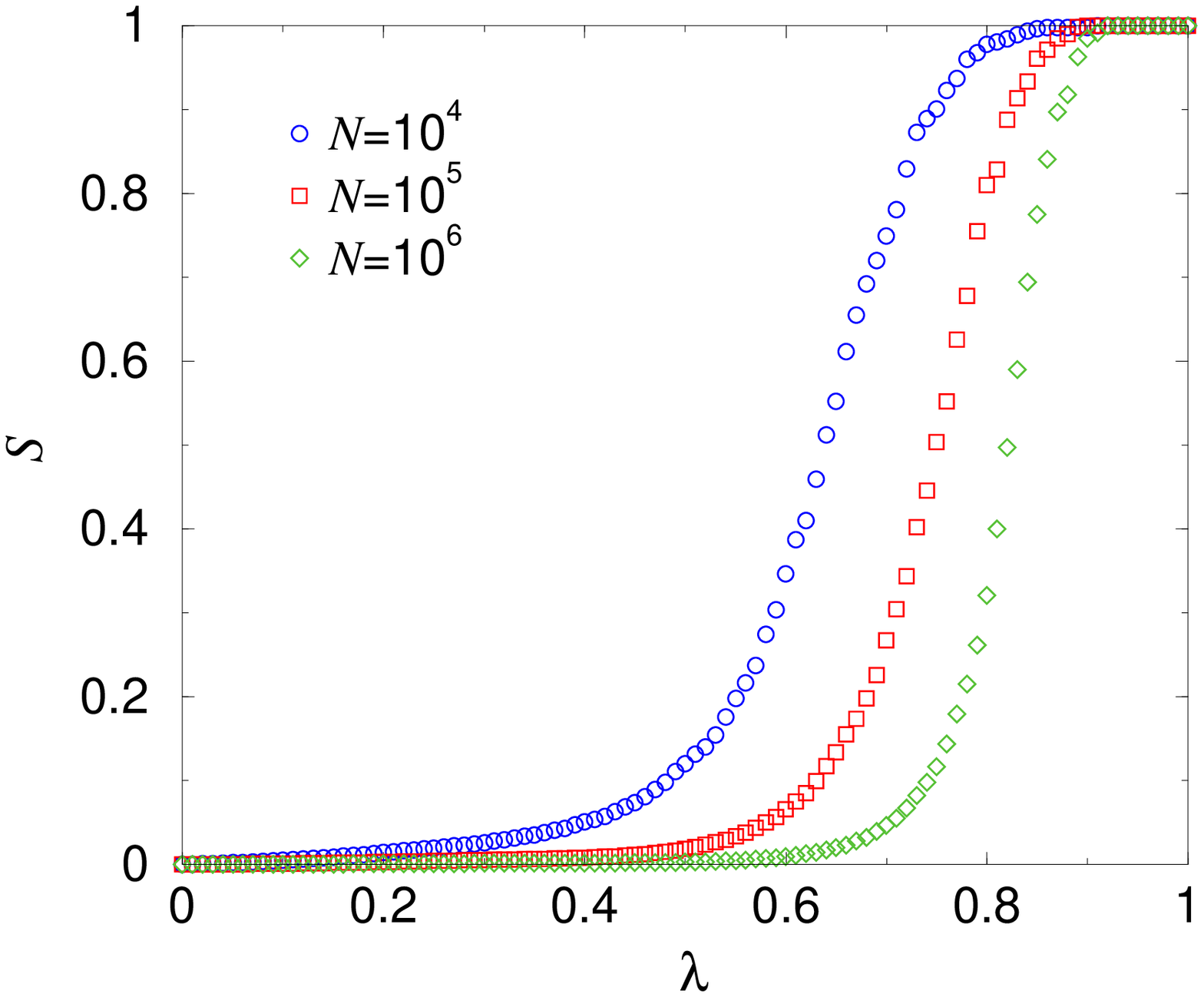}
\includegraphics[width=3in]{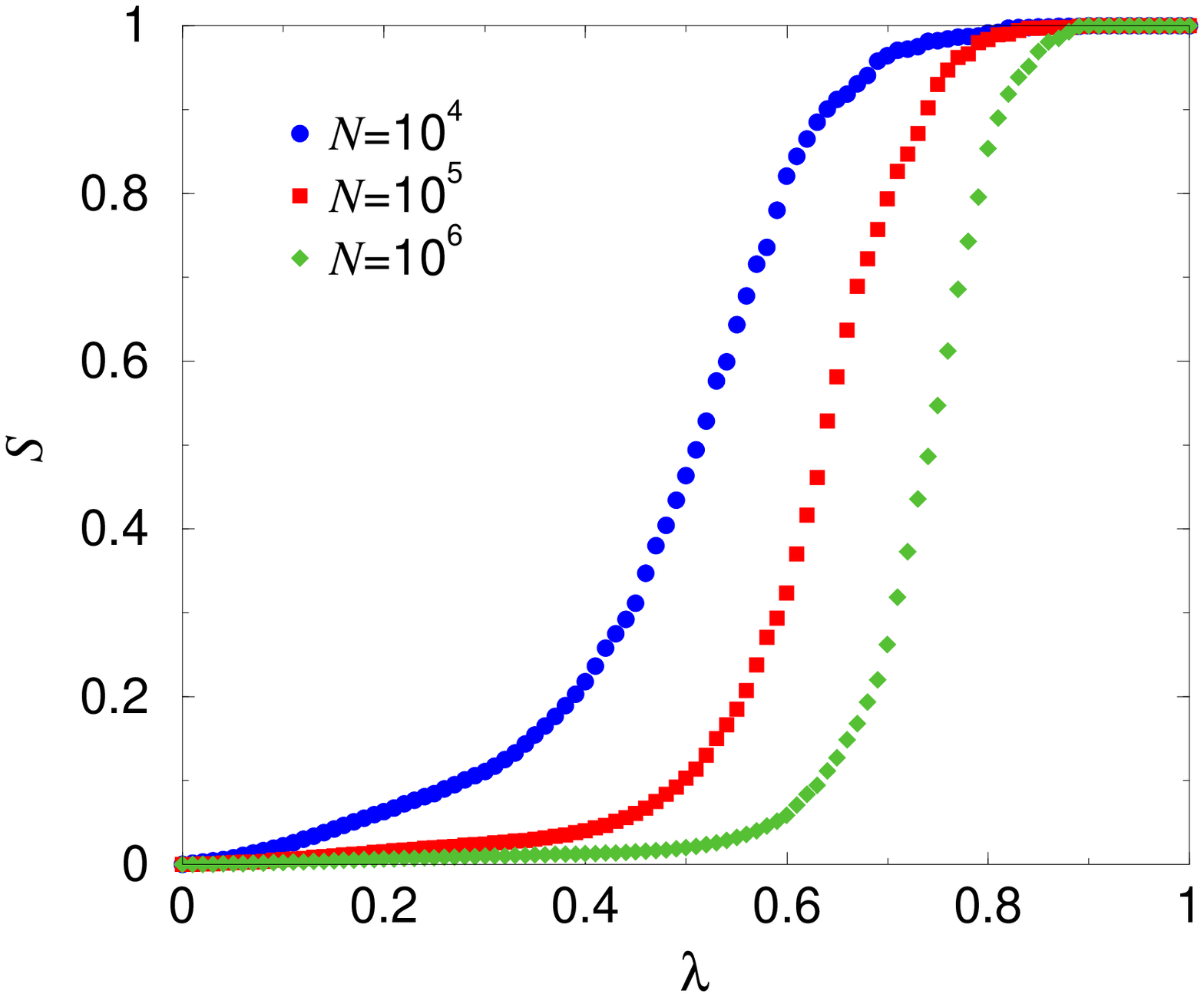}
\end{center}

\caption{Size of the giant component as a function of the bond
occupation probability $\lambda$ for the network generated with $m=3$. The
top figure (open symbols) corresponds to model A and the bottom
figure (filled symbols) to model B.}

\label{fig:SIR:1}
\end{figure}

The study of the SIR model in random SF networks confirmed the
epidemiological picture obtained for the SIS model in complex networks
with power-law connectivity distributions
\cite{lloydsir,moreno02}. For instance, it has been shown that the
effective epidemic threshold is inversely proportional to the
connectivity fluctuations $\fluck$ and hence it is vanishing in the
thermodynamic limit $N \to \infty$ for $\gamma \leq 3$. The high
heterogeneity of SF networks also causes that the relative incidence
of an outbreak strongly depends on the connectivity of the first
infected nodes \cite{moreno02}. This dependency on the initial seed
should not be confused with the dependency on the initial conditions
found previously since for the SIR model the differences in the
relative incidence of an epidemic outbreak is an intrinsic effect of
the large heterogeneity in the connectivity distribution of SF
networks and is not related to the presence of any topological
constrain.

We have made numerical simulations of the bond percolation problem on
top of the structured networks generated using the two variants of the
deactivation model, A and B. Once the graphs are generated each link
is removed with a probability $1-\lambda$. Then, the size of the giant
component of the resulting graph is computed. All the results reported
below were obtained taking an average over 10 graph realizations of
the deactivation model and 10 realizations of the link removal
procedure. In Fig. \ref{fig:SIR:1} we plot the size of the giant
component $S$ as a function of $\lambda$ for different graph sizes $N$
for models A and B, with $m=3$. Notice that in both cases the
qualitative picture is the same.

Between the limiting cases $S(0)=0$ and $S(1)=1$, there is an
intermediate range of $\lambda$ where $S$ goes from a value close to 0
to a value close to 1. The width of this interval decreases
appreciable with increasing $N$. Moreover, the point at which the
transition takes place systematically shifts to larger values
of $\lambda$ approaching 1. Since the deactivation model has
essentially a one dimensional topology we expect that in the large $N$
limit $S=0$ for any $\lambda<1$. This hypothesis can not be confirmed
by direct observation of Fig. \ref{fig:SIR:1} but it can be checked using
a finite size scaling analysis. When we plot $S$ as a function
$1-\lambda$ in a log-linear scale we observe equidistant shifts of the
curves after increasing $N$ from $10^4$ to $10^5$ and then
to $10^6$. This scenario corresponds to a scaling of the form
$S=f[(1-\lambda)N^\alpha]$, where $f(x)$ is a scaling function
that is independent of $N$ and $\alpha$ is a scaling exponent.

\begin{figure}[t]
\begin{center}
\includegraphics[width=3in]{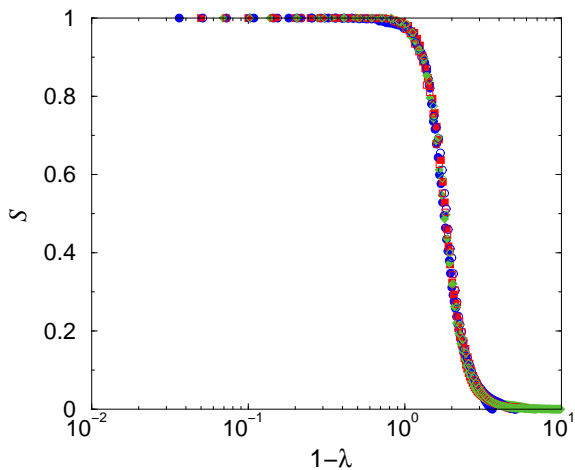}
\end{center}

\caption{Size of the giant component as a function of the rescaled
variable $(1-\lambda)N^\alpha$, with $\alpha=0.15$ and $\alpha=0.14$
for models A and B, respectively. The symbols are in correspondence
with those in Fig.\ \ref{fig:SIR:1}. All the curves of model A were
shifted by the same amount to the right to make evident the similarity
between the curves of both models.}

\label{fig:SIR:2}
\end{figure}

In Fig. \ref{fig:SIR:2} we plot $S$ versus $(1-\lambda)N^\alpha$ in a
log-linear scale using the value of $\alpha$ that gives the best data
collapse. This was achieved for $\alpha=0.15(1)$ and
$\alpha=0.14(1)$ for models A and B, respectively. Within the
numerical error, the exponent $\alpha$ is the same for both
models. Moreover the scaling function $f(x)$ is also unique, up to a
constant factor. These results point out that with regard to the bond
percolation problem, or the SIR model, the differences in the degree
distributions of model A and B are qualitatively irrelevant. 

\section{Discussion and conclusions}

We have study the dynamics of infectious diseases in structured
scale-free networks generated using the deactivation model introduced
by Klemm and Egu{\'\i}luz. By means of large scale numerical
simulations, we have shown that the existence or not of an epidemic
threshold for the SIS model depends on the initial density of infected
individuals when the connectivity fluctuations of the network are
unbounded. This effect is completely new and reflects the peculiar
topological nature of the networks generated with the deactivation
model algorithm. Guided by the analytical solution to the SIS model in
an ensemble of disconnected stars, we may summarize the SIS dynamics
as follows.

The linear topological nature of the networks provoke the dynamics of
the epidemic spreading to be almost purely diffusive, like in a
one-dimensional lattice. However, for model B the small value of
$\gamma$, that leads to the divergence of the connectivity
fluctuations, allows the existence of stars with very large
connectivities such that for any value of $\lambda$ the infection can
always get trapped in these stars. Hence, starting from homogeneous
initial conditions, the probability of hitting a star in which the
epidemic survives is not zero. This is not certainly the case if one
starts at a single infected node. In this case, there is no time for
the infection to spread through the network. It would be of further
interest to study this diffusion process in more details, for example,
by looking at the distribution of waiting times for each node in the
cicle infected $\rightarrow$ susceptible $\rightarrow$ infected. On
the other hand, when the connectivity fluctuations are finite (model
A), the probability of finding a star with degree $k > k_c(\lambda)$
is zero below $\lambda_c$ and thus the epidemic threshold is recovered
for all possible initial conditions.

Finally, the study of the SIR model confirms that the dominant factor in this
case is the chain structure of the network with star-like nodes
connected locally. The threshold for this epidemiological model
coincides, as it should, with the critical point of the corresponding
one-dimensional bond percolation problem that
gives an epidemic threshold $\lambda_c=1$. 


In summary, we have provided evidences that in this kind of networks
the dominant factor determining the behavior of both the SIS and the
SIR models is the linear topology of the network. The existence of high
degree nodes introduces new effects such as the dependency on the
initial conditions for the deactivation model B with $\gamma \le
3$. As a final remark, we should say that the results for the SIS and
the SIR models here reported cannot be directly extended to all
correlated networks. The peculiar topological features of the networks
generated with the deactivation model make them unique within the class of
correlated networks. Some recent works have begun to
\cite{marian3,cond182,cond450} address the SIS and the SIR models in
correlated networks establishing the general conditions for the
existence or not of an epidemic threshold. The extensive numerical
exploration of these models in such networks is still to do.

\begin{acknowledgement}
We thank A.\ Vespignani and R.\ Pastor-Satorras for helpful comments
and discussions. This work has been partially supported by the
European Commission - Fet Open project COSIN IST-2001-33555.
\end{acknowledgement}


\begin{thebibliography}{99}

\bibitem{barabasi02} A.-L. Barab{\'a}si, R. Albert,
Rev. Mod. Phys. {\bf 74}, 42 (2002).

\bibitem{dorogorev} S.~N. Dorogovtsev and J.~F.~F. Mendes,
Adv. Phys. {\bf 51}, 1079 (2002).

\bibitem{falou99} M. Faloutsos, P. Faloutsos, and C. Faloutsos,
Comput. Commun. Rev. {\bf 29}, 251 (1999).

\bibitem{alexei} R. Pastor-Satorras, A. V{\'a}zquez, and
A. Vespignani, Phys. Rev. Lett. {\bf 87}, 258701 (2001).

\bibitem{www99} R. Albert, H. Jeong, and A.-L. Barab{\'a}si, Nature
{\bf 401}, 130 (1999).

\bibitem{strog01} S.~H. Strogatz, Nature {\bf 410}, 268 (2001).

\bibitem{montoya02} J.~M. Montoya and R.~V. Sol{\'e},
J. Theor. Biol. {\bf 214}, 405 (2002).

\bibitem{wagner01} A. Wagner, Mol. Biol. Evol. {\bf 18}, 1283 {2001).

\bibitem{jeong01} H. Jeong, S. Mason, A.~L. Barab{\'a}si, and
Z.~N. Oltvai, Nature {\bf 411}, 41 (2001).

\bibitem{spsk} R.~V. Sol{\'e}, R. Pastor-Satorras, E. Smith, and
T. Kepler, Adv. Complex. Syst. {\bf 5}, 43 (2002).

\bibitem{vazquez} A. V{\'a}zquez, A. Flammini, A. Maritan, and
A. Vespignani, {\em Modelling of protein interaction networks}
(2001}, preprint cond-mat/0108043.

\bibitem{watts98} D.~J. Watts and S.~H. Strogatz, Nature {\bf 393},
440 (1998).

\bibitem {barab99} A.-L. Barab{\'a}si and R. Albert, Science {\bf
286}, 509 (1999).

\bibitem{newvir} M. E. J. Newman, S. Forrest, and J. Balthrop,
Phys. Rev. E {\bf 66}, 035101(R) (2002).

\bibitem{white} S. R. White, J. O. Kephart, and D. M. Chess, {\em
Computer Viruses: A Global Perspective}, in Proceedings of the 5th
Virus Bulletin International Conference, Boston, 1995.

\bibitem{murray} J. D. Murray, {\em Mathematical Biology} (Springer
Verlag, Berlin, 1993).

\bibitem{anders} R. M. Anderson and R. M. May, {\em Infectious
Diseases in Humans} (Oxford University Press, Oxford, 1992).

\bibitem{pv01a} R. Pastor-Satorras and A. Vespignani,
Phys. Rev. Lett. {\bf 86}, 3200 (2001).

\bibitem{pv01b} R. Pastor-Satorras and A. Vespignani, Phys. Rev. E
{\bf 63}, 066117 (2001).

\bibitem{barabasi00} R.~A. Albert, H. Jeong, and A.-L. Barab{\'a}si,
Nature {\bf 406}, 378 (2000).

\bibitem{newman00} D.~S. Callaway, M.~E.~J. Newman, S.~H. Strogatz,
and D.~J. Watts, Phys. Rev. Lett. {\bf 85}, 5468 (2000).

\bibitem{havlin01} R. Cohen, K. Erez, D. ben Avraham, and S. Havlin,
Phys. Rev. Lett. {\bf 86}, 3682 (2001).

\bibitem{berg02} J. Berg and M. L{\"a}ssig, {\em Correlated random networks}
  (2002), Preprint condmat/0205589.   

\bibitem{marian1} M. Bogu{\~n}{\'a} and R. Pastor-Satorras, {\em
Epidemic spreading in correlated complex networks}, (2002), Preprint
condmat/0205621.

\bibitem{marian3} M. Bogu{\~n}{\'a}, R. Pastor-Satorras, and
A. Vespignani, {\em Absence of epidemic threshold in scale-free
networks with connectivity correlations} (2002), Preprint cond-mat/0208163.

\bibitem{n02a} M.~E.~J. Newman, {\em Assortative mixing in networks},
(2002), Preprint cond-mat/0205405.

\bibitem{vw02} A. V\'azquez and M. Weigt, {\em Computational
complexity arising from degree correlations in networks} Preprint
cond-mat/0207035 (2002).

\bibitem{cond182} A. V\'azquez and Y. Moreno, {\em Resilience to
damage of graphs with degree correlations} Preprint cond-mat/0209182
(2002).

\bibitem{klemm02} K. Klemm and V.~M. Egu{\'\i}luz, Phys. Rev. E {\bf
65}, 036123 (2002).

\bibitem{cond183} A. V\'azquez M. Boguna, Y. Moreno,
R. Pastor-Satorras, and A. Vespignani, {\em Topology and correlations
in structured scale-free networks}, Preprint cond-mat/0209183 (2002).

\bibitem{structured} V.~M. Egu{\'\i}luz and K. Klemm,
Phys. Rev. Lett. {\bf 89}, 108701 (2002).

\bibitem{fspv} R. Pastor-Satorras, and A. Vespignani,
Phys. Rev. E {\bf 65}, 035108(R) (2002).

\bibitem{marro99} J. Marro and R. Dickman, {\em Nonequilibrium phase
transitions in lattice models} (Cambridge University Press, Cambridge, 1999).

\bibitem{newman02a} M. E. J. Newman, Phys. Rev. E {\bf 66}, 016128 (2002).

\bibitem{lloydsir} R.~M. May and A.~L. Lloyd, Phys. Rev. E {\bf 64},
066112 (2001).

\bibitem{moreno02} Y. Moreno, R. Pastor-Satorras, and A. Vespignani,
Eur. Phys. J. B {\bf 26}, 521 (2002).

\bibitem{cond450} M. E. J. Newman, {\em Mixing patterns in networks:
Empirical results and models} Preprint
cond-mat/0209450 (2002).

\end{thebibliography}
\end{document}